\documentclass[letter]{aa} 
\usepackage{graphicx}
\usepackage{natbib}
\usepackage{verbatim}
\usepackage{color}
\usepackage{multirow}

\newcommand{\tenpow}[1]{10$^{#1}$}
\newcommand{\scitenpow}[2]{#1$\times$10$^{#2}$}

\def\src{Pipe nebula}

\def\pa{P$_{\rm ang}$}

\def\avdpa{$\langle\Delta$P$_{\rm ang}\rangle$}

\def\pd{P$_{\rm deg}$}
\def\avpd{$\langle$P$_{\rm deg}\rangle$}

\def\vwei{$v_{\rm weight}$}
\def\avvwei{$\langle v_{\rm weight}\rangle$}
\def\vwei{$v_{\rm m1}$}
\def\avvwei{$\langle v_{\rm m1}\rangle$}

\def\av{$A_{v}$}

\def\dco{$^{12}$CO}
\def\tco{$^{13}$CO}

\def\cmt{cm$^{-3}$}
\def\kms{km~s$^{-1}$}

\def\deg{$^\circ$}

\def\msun{$M_{\odot}$}

\def\lsim{\mathrel{\rlap{\lower4pt\hbox{$\sim$}}\raise1pt\hbox{$<$}}} 
\def\gsim{\mathrel{\rlap{\lower4pt\hbox{$\sim$}}\raise1pt\hbox{$>$}}}

\begin{document}
   \title{Formation of dense structures induced by filament collisions.}
   \subtitle{Correlation of density, kinematics and magnetic field in the \src.}

   \author{P.\ Frau \inst{1,2}
	\and
	J.~M.\ Girart \inst{3}
	\and
	 F.~O.\ Alves \inst{4}
	\and
	G.\ A.\ P.\ Franco \inst{5}
	 \and
	 T.\ Onishi \inst{6}
	 \and
	 C.\ G.\ Rom\'an--Z\'u\~niga \inst{7}
	   }

\institute{
Instituto de Ciencia de Materiales de Madrid (CSIC), Sor Juana In\'es de la Cruz 3,
E-28049 Madrid, Spain
\\\email{pfrau@icmm.csic.es}
\and
Observatorio Astron\'omico Nacional, Alfonso XII 3, E-28014 Madrid, Spain
\and
Institut de Ci\`encies de l'Espai (CSIC--IEEC), Campus UAB, Facultat de Ci\`encies, C5p~2, E-08193 Bellaterra, Catalonia, Spain
\and
Max-Planck-Institut f\"ur extraterrestrische Physik, Giessenbachstr. 1, D-85748 Garching, Germany
\and
Departamento de F\'isica--ICEx--UFMG, Caixa Postal 702, 30.123-970 Belo Horizonte, Brazil
\and
Department of Physical Science, Osaka Prefecture University, Gakuen 1--1, Sakai, Osaka 599-8531, Japan
\and
Instituto de Astronom\'ia -- UNAM, Unidad Acad\'emica en Ensenada, Ensenada BC 22860, M\'exico
}

   \date{Received 28 October 2014 / Accepted }

  \abstract
   {The \src\ is a molecular cloud that lacks star formation feedback and has a relatively simple morphology and velocity structure. This makes it an ideal target to test cloud evolution through collisions.}
   {We aim at drawing a comprehensive picture of this relatively simple cloud to better understand the formation and evolution of molecular clouds on large scales.}
   {We use archival data to compare the optical polarization properties, the visual extinction, and the \tco\ velocities and linewidths of the entire cloud in order to identify trends among the observables.}
   {The \src\ can be roughly divided in two filaments with different orientations and gas velocity ranges: E--W at 2-4~\kms\ and N--S at 6--7~\kms. The two filaments overlap at the bowl, where the gas shows a velocity gradient spanning from 2 to 7~\kms. Compared to the rest of the \src, the bowl gas appears to be denser and exhibits larger linewidths. In addition, the polarization data at the bowl shows lower angular dispersion and higher polarization degree. Cores in the bowl tend to cluster in space and tend to follow the \tco\ velocity gradient. In the stem, cores tend to cluster in regions with properties similar to those of the bowl.}
   {The velocity pattern points to a collision between the filaments in the bowl region. The magnetic field seems to be compressed and strengthened in the shocked region. The proportional increase of density and magnetic field strength by a factor similar to the Alfv\'enic Mach number suggests a continuous shock at low Alfv\'enic Mach number under flux-freezing. Shocked regions seem to enhance the formation and clustering of dense cores. }

   \keywords{ISM: clouds -- ISM: evolution --  ISM: kinematics and dynamics -- ISM: magnetic fields -- ISM: dust, extinction -- ISM: individual objects (\src)}

   \maketitle

\section{Introduction}

The \src\ is an ideal test site to study early cloud formation. From this massive, nearby, elongated cloud (\tenpow{4}~\msun; 145~pc away; 18~pc$\times$3~pc in size) arises $^{12}$CO~1--0 emission that shows two filamentary structures: one elongated along the N--S direction at a $v_{\rm LSR}$ range of 6--7~\kms; and the other one along the E--W direction at 2--4~\kms\ \citep{onishi99}.
The magnetic field is strikingly uniform along the N--S direction with optical polarization degrees up to an outstanding 15\% value  (\citealp{alves08}, hereafter AFG08; \citealp{alves14}). Overall, the cloud seems to be magnetically dominated and turbulence seems to be sub-Alfv\'enic \citep{franco10}. Besides the simple structure, the cloud is at a very early stage of evolution with little to none star-formation feedback. Only 14 young stars have been detected, mostly in the B59 region at the west end, delivering a very low $\sim$0.06\%\ star formation efficiency \citep{forbrich09}. In addition, the cloud harbors more than 150 starless cores with very early chemistry that seem to be gravitationally unbound and pressure confined \citep{lada08,rathborne08,roman09,roman10,frau10,frau12}.

   \begin{figure*}[t]
   \centering
   \includegraphics[width=\textwidth,angle=0]{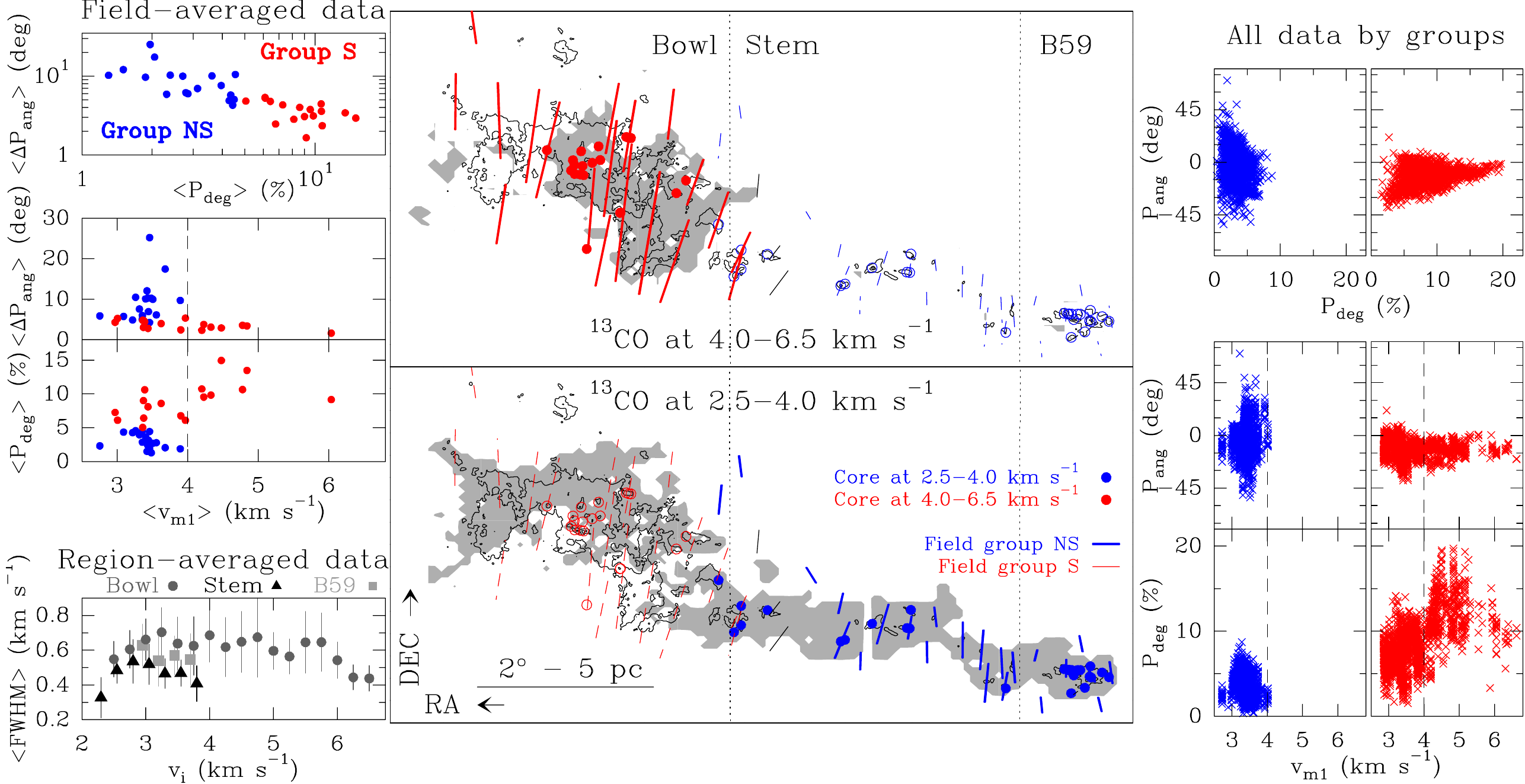}
   \caption{Velocity and polarization distribution of the \src.
   {\it Color code}: blue and red for field groups NS and S, respectively.
      {\bf Central maps}: 
        {\it Grayscale maps}: top and bottom panels show the regions with \tco\ \vwei\ in the $>$4~\kms\ and $<$4~\kms, respectively \citep[fitted to the map by][]{onishi99}. The N--S filament is evident in $^{12}$CO emission \citep{onishi99}.
       {\it Contours}: visual extinction map \citep{lombardi06}. Levels are 5.5 and 20~mag.
   {\it Segments}: average polarization segment on each observed field (AFG08; \citealp{franco10}). Length is proportional to \avpd. 
 {\it Circles}: positions of the embedded dense cores with extinction peaks $>10$~mag \citep{roman09,roman10}. 
   {\bf Left-hand side panels}: data averaged over the fields observed by AFG08. {\it Top panel}: \avdpa\ versus \avpd\ as in AFG08. {\it Central panels}: \avdpa\ and \avpd\ versus \avvwei, where m1 stands for the first-order moment of the \tco\ emission.
   {\it Bottom panel}: average linewidth of the \tco\ velocity components as a function of $v_{i}$ for the different cloud regions. $v_{i}$ is sampled in 0.25~\kms--wide bins. Error bars indicate 1-$\sigma$ dispersion.
   {\bf Right-hand side panels}: data for every pixel separated by groups. {\it Top panel}: \pa\ versus \pd. {\it Bottom panels}: \pa\ and \pd\ versus \vwei. 
   }
   \label{fig-vel}
    \end{figure*}

\section{Datasets and data process\label{sec-data}}

The three datasets used in this work are published in different articles. We limit ourselves to provide the basic information and we refer the reader to the original publications.

The dust extinction maps were derived by \citet{roman09,roman10} through the NICER infrared color excess technique \citep{lombardi01}. The angular resolution is 20\arcsec\ and the typical rms is $<$1~mag. 

The \tco~1--0 map was presented by \citet{onishi99} and has spectral and angular resolution of $\sim$0.1~\kms\ and 4\arcmin, respectively. The \tco\ emission is enclosed in the 5~mag contour of the \av\ map (Fig.~\ref{fig-vel}). Hence, \tco\ is a good tracer of the gas kinematics (Section~~\ref{ssec-resolution}).

The optical linear polarization data were published by AFG08 and \citet{franco10}. They observed 46 fields of 12\arcmin$\times$12\arcmin\ in the $R$-band, covering most of the \src. We used the $\sim$6,600 stars detected with $P/\sigma P\geq10$.

Our data processing followed two steps. First, the \tco\ datacube was converted to spectra on a pixel-by-pixel basis. A semi-automatic algorithm detected the number of peaks in each spectrum and carried out a Gaussian fit to derive the peak velocity, $v_{i}$, and the linewidth, FWHM$_{i}$, of the different components. For each pixel, we also derived the first-order moment of the \tco\ emission (\vwei). Second, we paired the position of each $R$-band star to its pixel in the \av\ and \tco\ map. Upon positional match, we composed a set of properties for each pixel: \av, \vwei, $v_{i}$, FWHM$_{i}$, and polarization angle (\pa) and degree (\pd).  The effects of the different angular resolutions are discussed in Section~\ref{ssec-resolution}.

   \begin{figure}
   \centering
   \includegraphics[width=\columnwidth,angle=0]{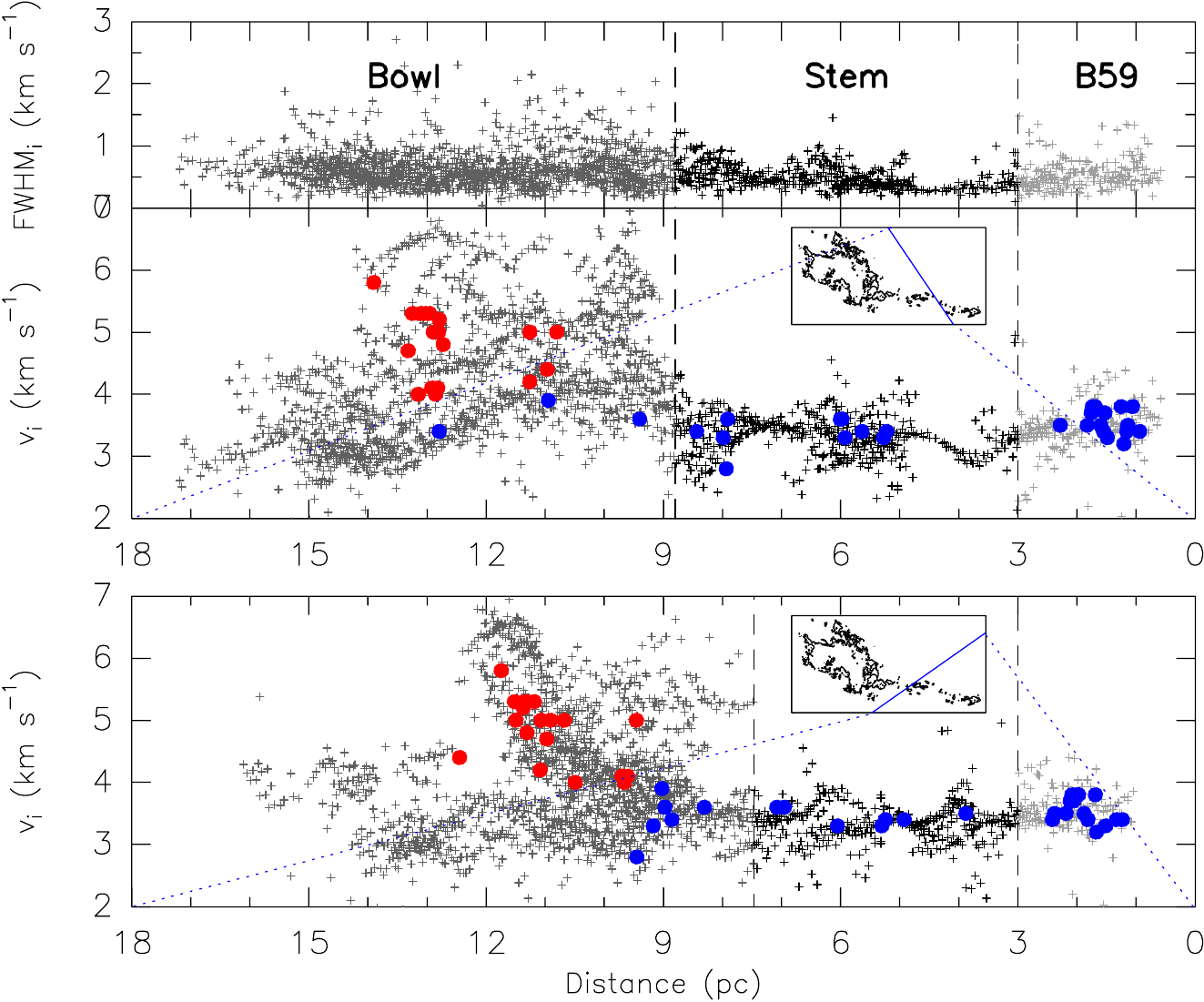}
   \vskip .1cm
   \includegraphics[width=.95\columnwidth,angle=0]{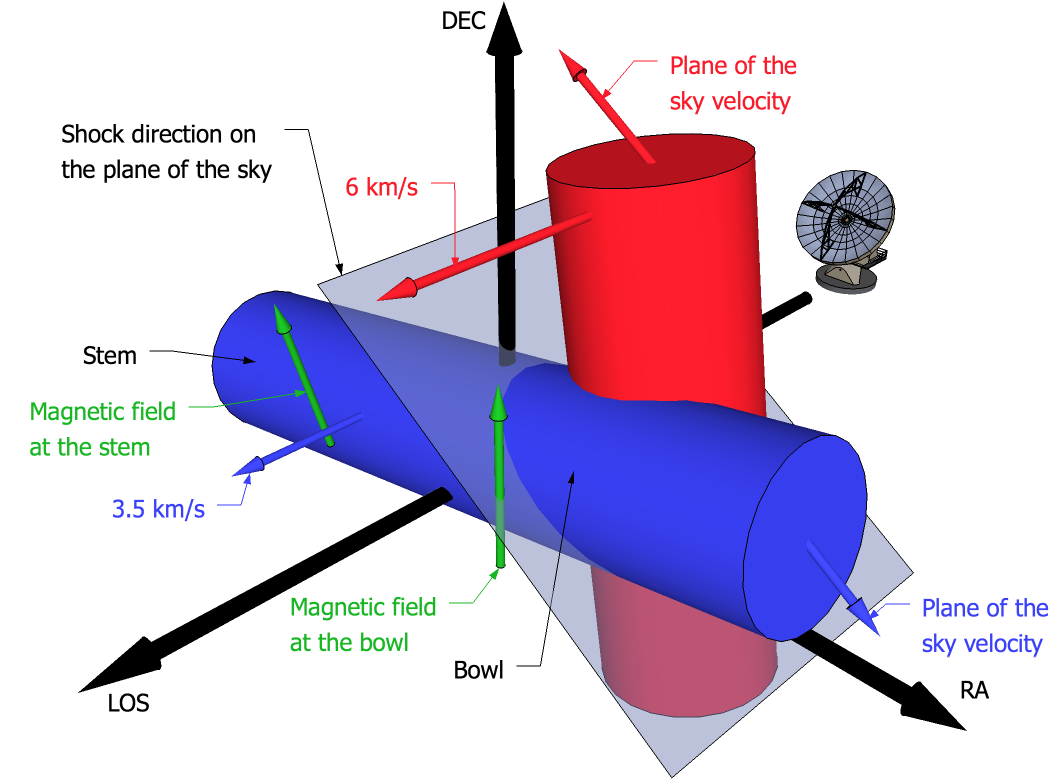}
 \caption{
 {\bf Top panels}: projections of the position-position-velocity cube in two perpendicular position-velocity planes. The orientation of the projection planes is indicated as blue lines in the \av\ map subpanels. {\it Bottom panel}: velocity as a function of position. The projection plane is at an angle of 124\deg\ (shock direction: projection on the plane of the sky of the velocity difference of the filaments). {\it Top panels}: $v_{i}$ and FWHM$_{i}$ as a function of position. The projection plane is at an angle of 34\deg\ (perpendicular to the shock direction). {\it Circles}: dense cores as in Fig.~\ref{fig-vel}.
{\bf Cartoon}: proposed scenario. 
 The antenna represents the observer. Blue and red cylinders represent the E--W and N--S filaments, respectively. The LOS velocities of the filaments are shown as arrows and the stem and bowl regions are labeled. The translucent plane shows the shock direction on the plane of the sky extended along the line-of-sight direction (blue line on bottom panel above). The velocities of the filaments along the shock are shown as arrows. Green arrows represent the tentative direction of the magnetic field (Section~\ref{ssec-mag-field}). }
   \label{fig-ppv}
    \end{figure}

\section{Results and analysis\label{sec-results}}

\subsection{Dual large-scale velocity structure}

Figure~\ref{fig-vel} outlines the main velocity structures of the \src\ derived from the \tco\ spectra. The cloud can be separated in two different filaments with narrow velocity ranges: one oriented E--W at $<$4~\kms\ and another one oriented N--S at $>$4~\kms\ \citep[this filament is better seen in the \dco\ map: see][]{onishi99}. The spectra from the E--W filament typically have one velocity component, two in some cases. Instead, the spectra from the N--S filament show a richer velocity structure with up to four components. Both structures overlap along the line-of-sight (LOS) at the bowl (Fig.~\ref{fig-vel}). The velocities at the overlapping region cover the entire range between those of the two filaments. This transition is illustrated in Fig.~\ref{fig-ppv} through two orthogonal projections of the position-position-velocity cube \citep[see][]{hacar13} that show distinct patterns (Section~\ref{ssec-collision}).

We define two regions: the non-shocked gas --group NS-- that roughly includes B59 and the stem (\avpd$\la$5\%) and the shocked gas --group S-- that roughly coincides with the bowl (\avpd$\ga$5\%). These groups appear color-coded on Fig.~\ref{fig-vel} and also have distinct properties in polarimetry, density, and turbulence (Section~\ref{ssec-resPol}).

\subsection{Bimodal trends in velocity, polarization, and linewidth\label{ssec-resPol}}

The top left panel on Fig.~\ref{fig-vel} replicates Fig.~2 by \citet{alves08}. They report a systematic anti-correlation between \avdpa\ and \avpd. The color-coded NS and S groups appear clearly differentiated: S fields have high \avpd\ and low \avdpa, oppositely to NS fields. This differentiation goes beyond field averages and holds at a single pixel level (top right panels). 

The central left panels on Fig.~\ref{fig-vel} show the distinct distribution of \avdpa\ and \avpd\ as a function of \avvwei. The bottom right panels of Fig.~\ref{fig-vel} show similar plots but for all the pixels. These panels show that the clear polarization differences between group~S and group~NS are mostly independent of the velocity of the gas, thus suggesting different scenarios for the two regions. Group~S gas at 4--5~\kms\ shows the largest polarization degree.

The bottom left panel of Fig.~\ref{fig-vel} and top panel of Fig.~\ref{fig-ppv} show that the gas in the bowl has a larger linewidth than in B59 and the stem. Strikingly, the star-forming B59 complex has a smaller linewidth than group~S. This suggests that the Bowl gas is more disturbed than that of B59.

\section{Discussion\label{sec-disc}}

\subsection{Data suitability and spatial resolution\label{ssec-resolution}}

The polarization data trace material with \av  $\la$15~mag that includes 99.1\% of the \av\ maps pixels, and \citet{franco10} show that \pd\ and \av\ grow roughly proportionally. Hence, optical data are good tracers of the magnetic field of the cloud. The NICER \av\ maps are very accurate: a comparison to 1.2~mm emission maps of dense cores shows a perfect within-the-errors agreement for up to 45~mag \citep{frauthesis}. The spatial resolution of the \av\ map is $\sim$0.015~pc. This is significantly smaller than the typical 0.1~pc diameter of dense cores, and thus, the map resolution is more than enough for the large scales. \tco\ is a good tracer of the dense gas (Fig.~\ref{fig-vel}). The \tco\ pixel size is $\sim$0.17~pc, comparable to a dense core. With a mean \tco\ linewidth of 0.6$\pm$0.3~\kms, the pixel crossing time is $\sim$0.28~Myr. The dense core formation timescale is 0.5--1~Myr \citep{bergin07}, and thus, it seems unlikely that any velocity substructure would form. Therefore, the \tco\ map should suffice to reflect all significant variations in the gas velocity.

\subsection{Interaction of gas filaments and core formation\label{ssec-collision}}

The two filaments overlap at the bowl and the gas velocities cover the entire range between the original velocities (Fig.~\ref{fig-ppv}). This fact points to an interaction, possibly a collision, as reported toward denser and more massive IRDCs \citep{duartecabral11,henshaw13,nakamura14}. The sound speed at 12~K is $c_{s}$=0.27~\kms\ and the pre-shock Alfv\'en speed\footnote{Assuming a magnetic field on the plane of the sky with $B_{\rm stem}$=30~$\mu$G, and a density of \scitenpow{3}{3}~\cmt\  (AFG08).} is $v_{\rm A}$=0.78~\kms. The magnetosonic speed $v_{m} = \sqrt{c_{s}^{2}+v_{A}^{2}}$ is 0.83~\kms, smaller than the velocity separation of the filaments. Therefore, an MHD wave is slower than the velocity difference between the filaments and a shock would indeed form. Hence, in the \src\ it is possible to compare the evolution from non-shocked (stem) to shocked (bowl) gas.

The two orthogonal projections of the position-position-velocity cube show that LOS velocities are anisotropic (Fig.~\ref{fig-ppv}). On the one hand, the bottom panel shows highly ordered velocities likely to be the velocity gradient generated along the direction of the collision. On the other hand, the middle panel shows a large velocity dispersion likely to be the projection across the shock. This identified direction for the plausible shock is the projection on the plane of the sky of the velocity difference between the filaments. In addition, for the collision to happen, the N--S filament must be closer to us because it is going away faster than the E--W filament. The total shock velocity would be a combination of two components: (a) the line-of-sight velocity and (b) the plane-of-the-sky velocity whose direction is deduced from the projections of the position-position-velocity cube. Figure~\ref{fig-ppv} shows a cartoon of the proposed scenario. 

Figure~\ref{fig-vel} shows that linewidths increase {\it only} at the bowl {\it and} for the velocities between the two filaments (roughly 4 to 6~\kms). This points to a more turbulent region where the interaction takes place. The interaction would settle the gas at the observed intermediate velocities and result in sub- and trans-Alfv\'enic gas motions \citep{franco10}. The accumulation of material would cause the increase in density and would lead to the formation of the bowl. In fact, the average column density of 6.0$\pm$1.8~mag at the bowl is twice as high as the average 3.0$\pm$1.3~mag at the stem. This increase reflects an increase in volume density due to the plausible filament collision.

The bowl is the region with the largest column density and the most fertile in number of dense cores (Fig.~\ref{fig-vel}). B59, the other region rich in dense cores, is also proposed to be formed through an external shock that triggered star formation \citep{peretto12}. Cores have velocities in good agreement with gas velocities. At the bowl, cores cover the entire velocity range between the filaments, follow the velocity gradient along the shock, and tend to cluster at the larger velocities (bottom panel in Fig.~\ref{fig-ppv}). This combined structure of cores and gas is the ``molecular ring'' reported by \citet{muench07}. There are other clusters of cores, one in B59 and two in the stem (middle panel in Fig.~\ref{fig-ppv}). The gas surrounding these clusters shows similar properties: a larger dispersion of velocities, and increases in linewidth, \tco\ brightness, and column density. These properties resemble those of the shocked bowl but with smaller values and at a smaller spatial scale. In general, it seems that in the slowed, shocked gas the higher density and sub-Alfv\'enic motions enhance the creation of cores.

\subsection{Magnetic field compression and polarization increase\label{ssec-mag-field}}

The high \pd\ values --close to the theoretical maximum \citep{whittet03}-- suggest that the magnetic field direction is close to the plane of the sky AFG08. The higher \pd\ values are detected towards the bowl, where the two filaments interact. The shock can have reoriented the magnetic field placing it closer to the plane-of-the-sky (Fig.~\ref{fig-ppv}). This scenario could explain the variation of \pd. Moreover, the peak of \pd\ is at a velocity of $\sim$4.5~\kms, the intermediate velocity between the filaments where the shock would have the strongest effects.

The material traced by the optical polarization is dominated by magnetic field over gravity and turbulence \citep{franco10}. This low-density gas is in flux-freezing conditions, this is, matter and magnetic field are tied. As expected in this regime, the large-scale gas compression that causes the column density to double also doubles the magnetic field strength: from 30~$\mu$G at the stem to 65~$\mu$G at the bowl (AFG08). This implies that the magnetic energy density is four times higher in the bowl than in the stem ($E_{\rm mag}^{\rm stem}$$\simeq$\scitenpow{3.6}{-11}~erg~\cmt\ and $E_{\rm mag}^{\rm bowl}$$\simeq$\scitenpow{1.7}{-10}~erg~\cmt). Similarly, the larger line widths in the bowl than in the stem (0.6$\pm$0.3 and 0.48$\pm$0.19~\kms, respectively) give a turbulent kinetic energy density about 40\% higher in the bowl ($E_{\rm kin}^{\rm bowl}$$\simeq$\scitenpow{2.3}{-11}~erg~\cmt). This generates an increase of the surface pressure on dense cores, reported to be $\sim$60\% higher in the bowl with respect to the stem \citep{lada08}. The energy increase can be explained by the plausible shock scenario: assuming an average volume density of \scitenpow{3}{3}~\cmt\ (AFG08) and that the shock velocity is the velocity difference between the two filaments, 2.5~\kms, then the shock energy density is $E_{\rm shock}$$\simeq$\scitenpow{3.7}{-10}~erg~\cmt, twice the energy increase from the stem to the bowl. Consequently, the kinetic energy of the collision could increase the measured turbulence levels, and then drag and compress the magnetic field, transferring and storing some of the kinetic energy into magnetic field energy. 

The Alfv\'enic Mach number $M_{A}=v_{\rm shock}/v_{A}$ of the plausible shock is $\sim$3.0. Simulations of low $M_{A}$ (highly magnetized) collisions predict the formation of shocks \citep{lesaffre13}. Indeed, models with low Alfv\'enic Mach numbers ($M_{A}$$\sim$4--5) drive more energy into compressing the magnetic field \citep{pon12}. In particular, for clouds of similar temperature and density to the \src\ colliding at 2--3~\kms, the density and magnetic field strength increase by a factor similar to $M_{\rm A}$. Moreover, 20--50\% of the shock energy is stored in the magnetic field. These predictions match our results at the \src, also if taking the factor of 2 error in the measured quantities into account. This provides a plausible theoretical framework that reproduces our observations. The majority of the remaining energy is expected to be radiated away through mid-J CO lines that can be used to further test this scenario.


\section{Conclusions\label{sec-concl}}

We combined visual extinction, optical polarimetry, and \tco\ data of the \src. Here, we summarize the conclusions.

\begin{enumerate}

\item The velocity patterns of the gas suggest that the N--S and E--W filaments are colliding along the NW-SE direction at the bowl, where turbulence increases.

\item There is a good correlation between polarization properties and velocity of the gas, suggesting that magnetization increases at the interaction site. The column density and magnetic field strength double with respect to the non-shocked gas, pointing to a highly magnetized, continuous collision under flux-freezing.

\item Cores in the bowl tend to cluster in space and tend to follow the \tco\ velocity gradient. Dense cores tend to form in regions similar to the bowl where the cloud gas has large velocity dispersion, linewidths, column density, and \tco\ emission. It seems that in the slowed, shocked gas the higher density and sub-Alfv\'enic motions enhance the creation of cores.

\end{enumerate}


\begin{acknowledgements}

We thank Daniele Galli and Andy Pon for useful comments and suggestions that are greatly appreciated. P.F. is supported by the CONSOLIDER project CSD2009-00038, Spain. P.F. and J.M.G. are supported by the MINECO project AYA2011-30228-C03-02, Spain. G.A.P.F. is supported by the CNPq agency, Brazil. CRZ is supported by program CONACYT CB-2010 152160, M\'exico.

\end{acknowledgements}


\end{document}